\documentclass{epl}

\usepackage{amssymb}

\newcommand \be {\begin{equation}}
\newcommand \ee {\end{equation}}
\newcommand \bea {\begin{eqnarray}}
\newcommand \eea {\end{eqnarray}}

\title{Real space analysis of inherent structures}
\author{Eric Bertin\inst{1,2}}
\institute{
  \inst{1} Department of Theoretical Physics, University of Geneva,
CH-1211 Geneva 4,\\
Switzerland\\
  \inst{2} SPEC, CEA Saclay, F-91191 Gif-sur-Yvette Cedex, France}


\begin{document}

\maketitle

\begin{abstract}
We study a generalization of the one-dimensional disordered Potts model,
which exhibits glassy properties at low temperature.
The real space properties of inherent structures visited dynamically are
analyzed through a decomposition into domains over which the energy is
minimized. The size of these domains is distributed exponentially, defining a
characteristic length scale which grows in equilibrium when lowering
temperature, as well as in the aging regime at a given temperature.
In the low temperature limit, this length can be interpreted
as the distance between `excited' domains within the inherent structures.
\end{abstract}

A significant part of our understanding of the low temperature behavior of
disordered systems comes from the landscape picture, which was first
introduced as a free-energy landscape, in the context of mean-field
spin-glasses \cite{TAP,Bray,Kurchan93}
--yet more qualitative landscape ideas for
structural glasses appeared earlier \cite{Goldstein}.
A complementary approach, more suitable for numerical simulations,
is to consider the potential energy (instead of free-energy) landscape,
which relies on the notion of inherent structure (IS). These IS are defined
as local minima of the potential energy in phase space,
i.e.~they are stable against a set of elementary transitions defined by
the dynamical rules.
This notion has been shown to be very useful for the understanding of
glassy dynamics, both in disordered \cite{CrisRitort00,Sellitto,CrisRitort02}
and non disordered systems \cite{Stillinger,Heuer}.

On the other hand, many efforts have also been devoted to the understanding
of the real space properties of disordered systems. This includes
in particular the structure of low energy excitations over the ground state
\cite{McMillan,Krzakala},
as well as the idenfication of a correlation length through the definition
of a four-point correlation function, which compares the dynamical state
of two copies of the system having the same disorder, but independent
thermal histories \cite{Marin96,Huse91,Rieger95,Rieger96,Marin00}.
By definition, the length scale deduced from this four-point correlation
function characterizes the current configuration visited dynamically by
the system.
Yet, since IS are believed to play an important r\^ole in the dynamics of
disordered systems, one may wonder whether there also exists a length scale
characterizing the underlying IS.
A few examples of spatial analysis of IS have been known for long --see
e.g.~\cite{DerridaGardner} for the disordered Ising chain-- but this analysis
was considered as a technical procedure to compute the configurational
entropy rather than a way to introduce a characteristic length scale.
Besides, the issue of characteristic length scales for IS was addressed
recently in the context of kinetically constrained models (models with no
interactions, but with kinetic constraints), where IS are found to differ
from the ground state only by a set of local defects \cite{Berthier-JPG}.
Still, in systems with more complex hamiltonians than kinetically constrained
models, including for instance disordered interactions between sites,
the spatial structure of IS essentially remains an open issue.

In this Letter, we show within the context of a simple disordered model that
IS exhibit a non trivial real space structure.
A numerical algorithm allows us to decompose explicitely an IS into
essentially non-overlapping `regions' upon which the IS identifies with an
absolute energy minimum. The size of these regions is found to be distributed
exponentially, leading unambiguously to a characteristic length scale $\xi^*$,
which grows in the aging regime before saturating to an equilibrium value that
increases when lowering temperature. An interpretation in terms of
small excited domains in the low temperature limit is also proposed.

\begin{figure}
\includegraphics[width=65mm,clip]{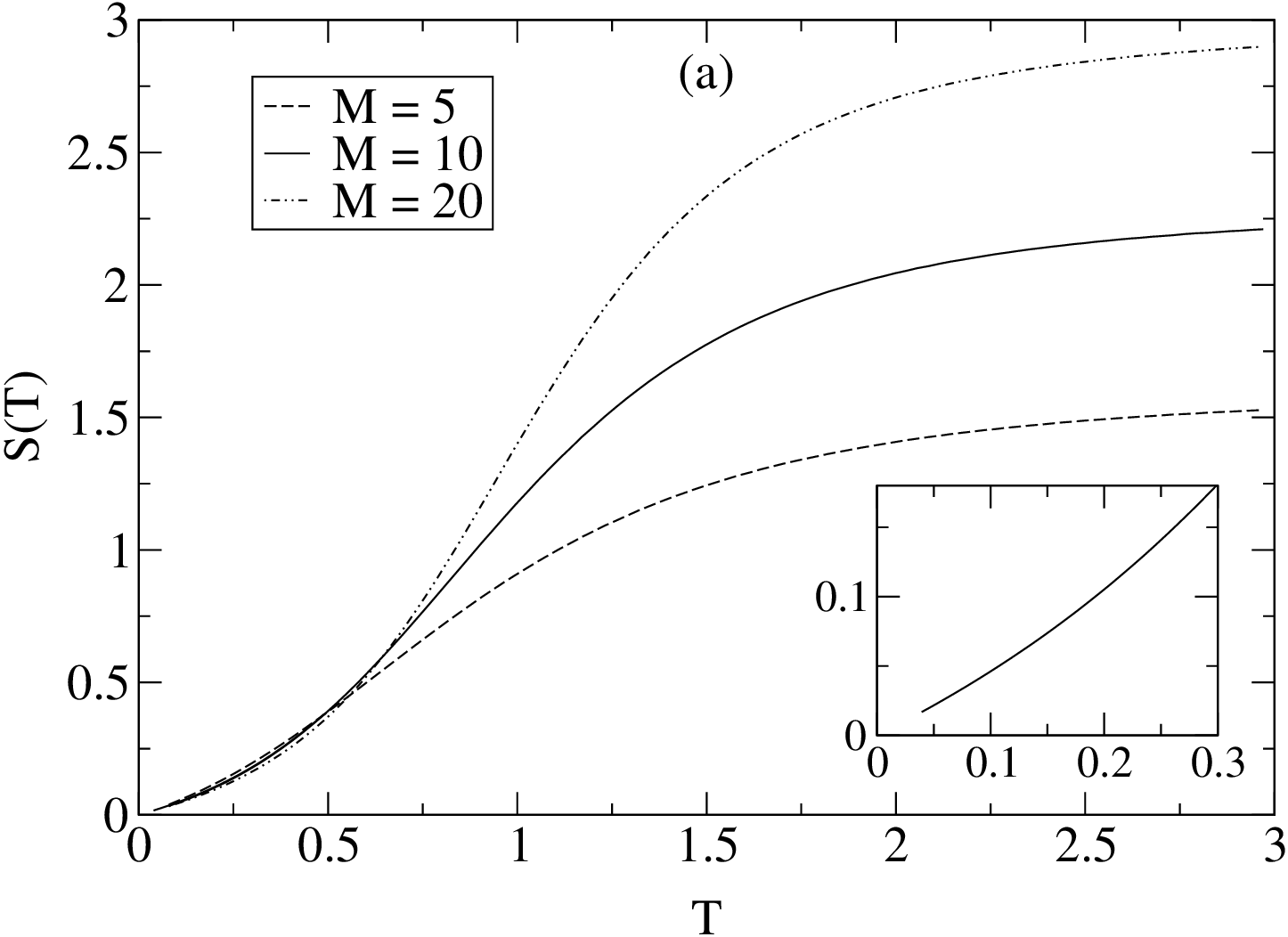} \hfill
\includegraphics[width=65mm,clip]{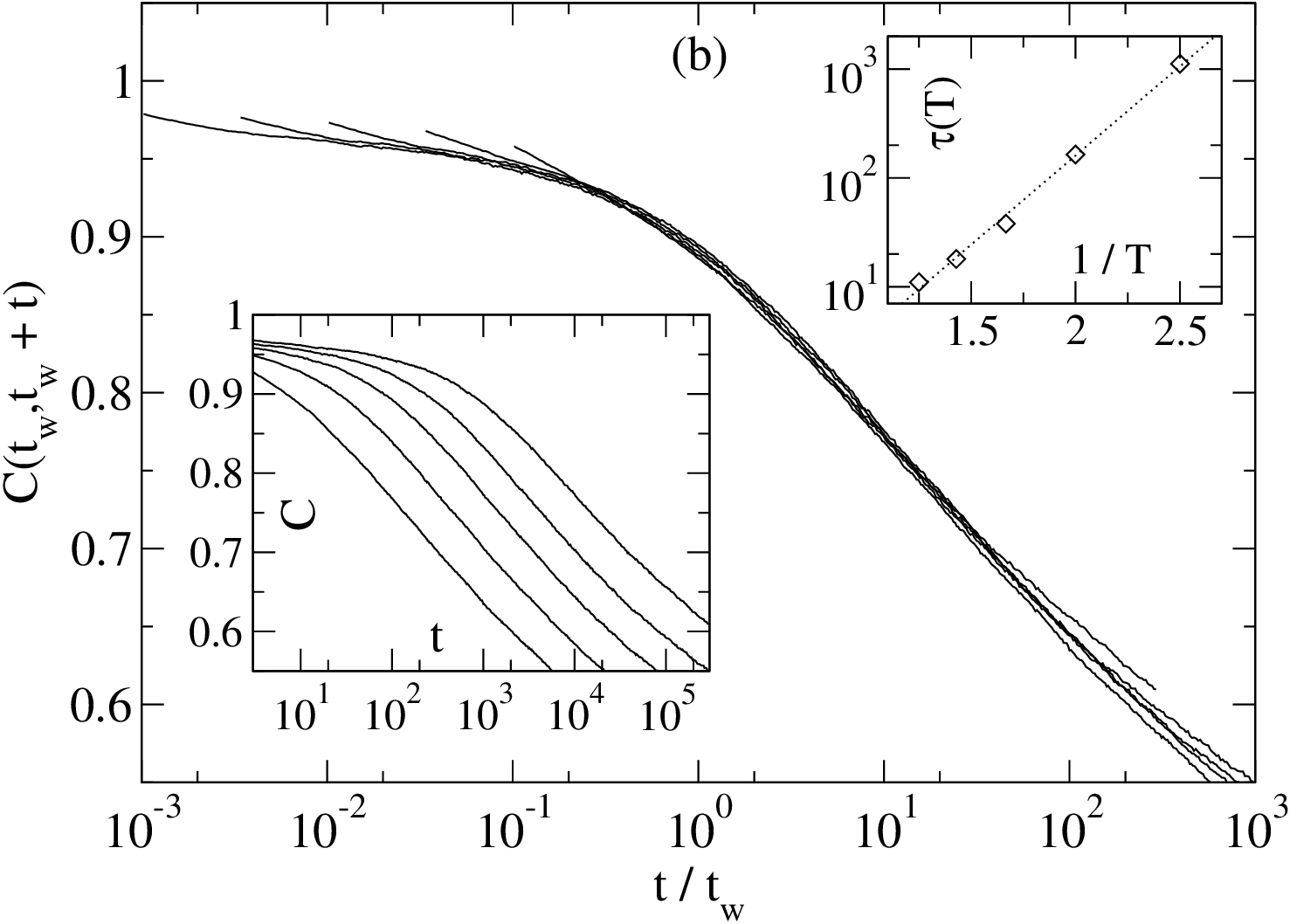}
\caption{\sl (a) Equilibrium entropy $S(T)$ as a function temperature $T$;
no static transition is found at finite temperature. Inset: zoom on small
temperatures for $M=10$. (b) Correlation function $C(t_w,t_w+t)$ for $T=0.2$,
$M=10$ and $t_w = 10$, $30$, $100$, $300$ and $1000$ as a function of the
ratio $t/t_w$, showing a simple aging behavior. Left inset: same data
plotted as a function of $t$. Right inset: equilibrium relaxation time
$\tau(T)$, showing Arrhenius behavior $\ln \tau \sim 1/T$.
}
\label{fig-entrop}
\end{figure}

\section{Model and thermodynamics} The model we introduce below is aimed at
being a simplified coarse-grained description of disordered systems.
For simplicity, only the one-dimensional case is considered in the following.
We assume that space is divided into cells, labeled by an index $i=1\ldots N$,
 and that the internal state of the system within each cell can be described
by a variable $q_i$ taking integer values between $1$ and $M$.
The set of variables $\{q_i\}$ is denoted by $q$ in the following.
To account for the interactions between neighbouring cells, we introduce
an interaction energy $V_{i,i+1}(q_i,q_{i+1})$, which a priori takes
different values for each couple of variables $(q_i,q_{i+1})$, and for
each link $(i,i+1)$. To be more specific, the interaction terms
$V_{i,i+1}(q_i,q_{i+1})$ are independent quenched random variables drawn
from a distribution $\rho(V)$ for each link and for each value of
$(q_i,q_{i+1})$; note that $V_{i,i+1}(q,q') \ne V_{i,i+1}(q',q)$.
The hamiltonian is defined by $H=-\sum_{i=1}^N V_{i,i+1}(q_i,q_{i+1})$,
with periodic boundary conditions.
This model can be thought of as a generalization of the disordered Potts
model. In the latter one assumes that $V_{i,i+1}(q_i,q_{i+1})$ can take only
two distinct values, one for $q_i=q_{i+1}$ and the other for
$q_i \neq q_{i+1}$.
In the present model, $V_{i,i+1}(q_i,q_{i+1})$ takes $M^2$
distinct values, as in the physical picture described above, there is no
reason why the interaction energies should be equal for different values
of $(q_i,q_{i+1})$.

In the following, we restrict ourselves to the case of
an exponential distribution
$\rho(V)=V_0^{-1} e^{-V/V_0}$; $V_0$ is set to unity in the following.
We have studied the thermodynamic properties of this model, namely the
free energy density $F(T)$ and the entropy density $S(T)$.
The canonical partition function is obtained by computing numerically the
product of transfer matrices, which allows to reach system sizes as large
as $N=10^5$.
The entropy $S(T)$ is plotted on Fig.~\ref{fig-entrop}(a) for different
values of $M$; one sees that the entropy remains regular when
$T \to 0$.

\section{Aging properties} Turning to dynamics, one needs first to define
the kinetic rules according to which the system evolves. We use a one-site
Glauber dynamics at temperature $T$, with a transition rate
associated to site $i$,
$W_i(q \to q') = \tau_0^{-1}/[1+\exp(\Delta U_i/T)]$,
where $\Delta U_i = U_i(q')-U_i(q)$.
The microscopic time scale $\tau_0$ is taken as the time unit
in the following. The local energy $U_i(q)$
is the sum of the interaction terms with the neighbors of site $i$,
$U_i(q) = V_{i-1,i}(q_{i-1},q_i)+V_{i,i+1}(q_i,q_{i+1})$.

\begin{figure}
\includegraphics[width=65mm,clip]{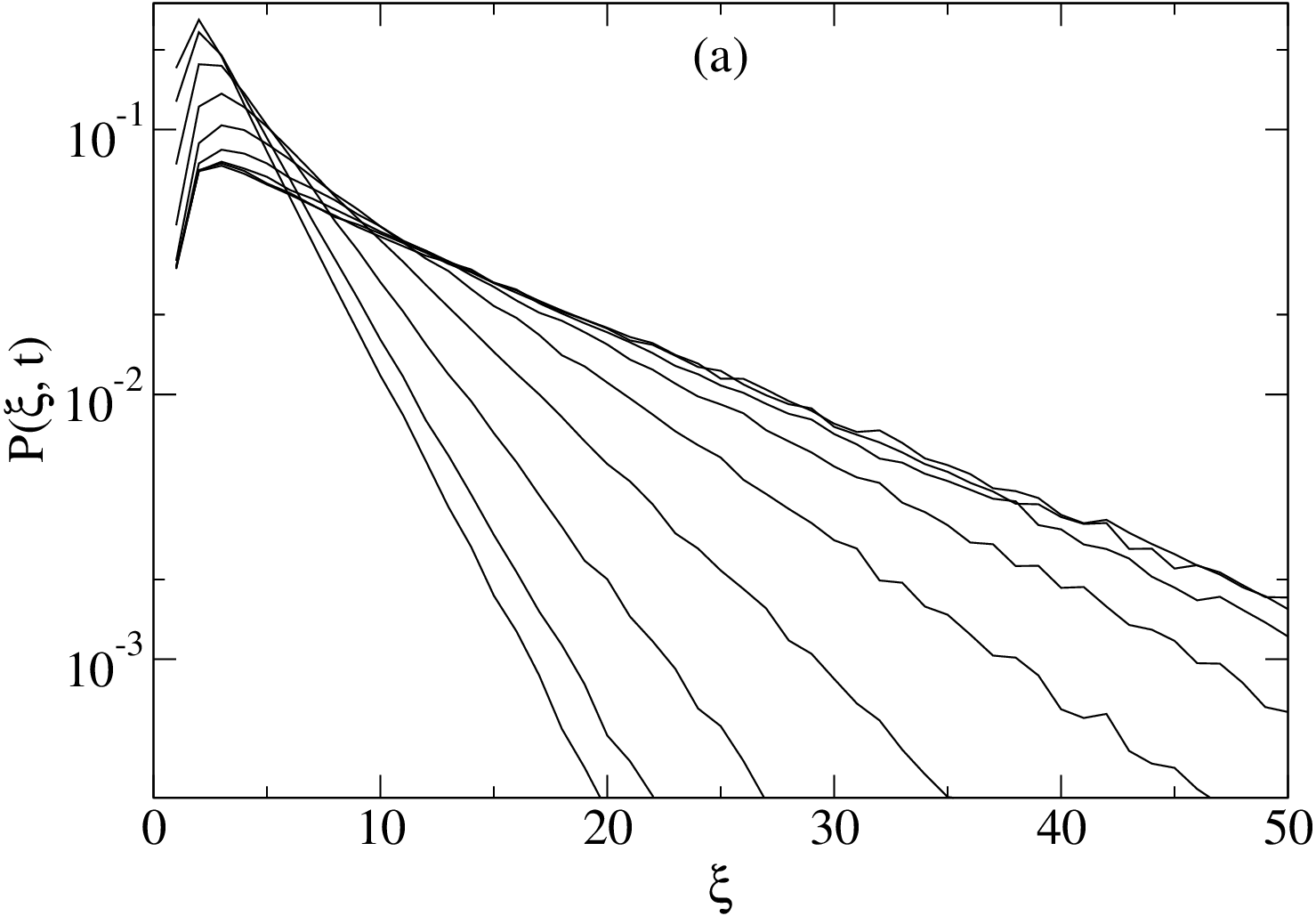} \hfill
\includegraphics[width=65mm,clip]{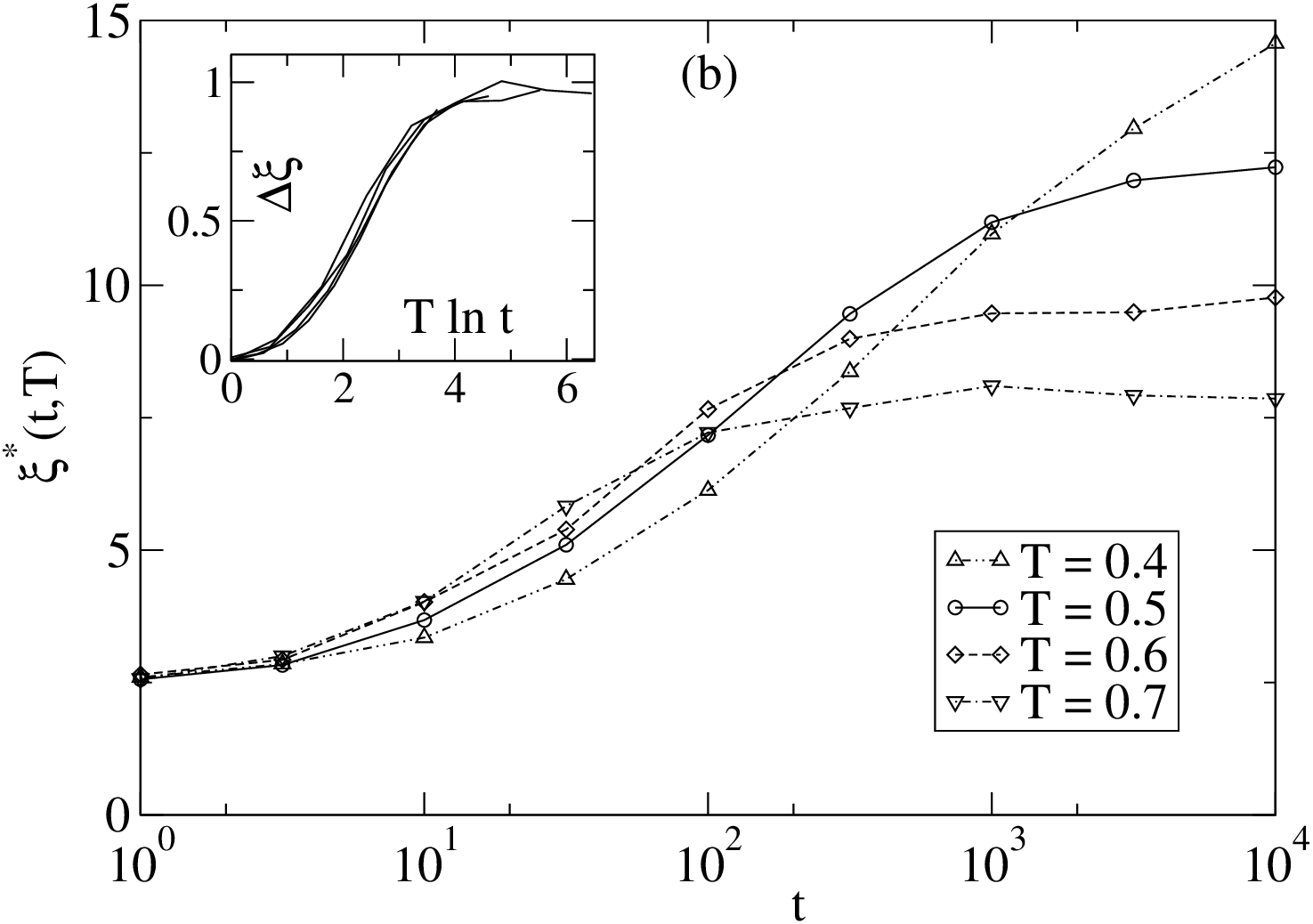}
\caption{\sl (a) Distribution $P(\xi,t)$ of the size $\xi$ of the regions
upon which energy is minimized, for $T=0.5$, $M=10$ and times ranging from
$t=1$ to $t=10^4$. The distribution is exponential, except for
$\xi \lesssim 4$. (b) Characteristic scale $\xi^*(t,T)$ of $P(\xi,t)$ as a
function of time, for different temperatures ($M=10$).
Inset: $\Delta\xi \equiv (\xi^*(t,T)-\xi_0)/(\xi_{eq}(T)-\xi_0)$ vs.~$T\ln t$,
showing a rather good collapse; $\xi_0$ is the initial value of $\xi^*$,
which does not depend on $T$, and $\xi_{eq}(T)$ is the equilibrium length.
}
\label{fig-Pell}
\end{figure}

So as to give evidence for the aging dynamics that takes place at low
temperature before the system equilibrates, we introduce a simple correlation
function $C(t_w,t_w+t)$ defined as the average overlap between configurations
occupied at time $t_w$ and $t_w+t$:

\be
C(t_w,t_w+t) = \left< \frac{1}{N}\, \sum_{i=1}^N \delta_{q_i(t_w),q_i(t_w+t)}
\right>
\ee
where the brackets $\langle \dots \rangle$ denote an average over the
thermal histories.
Numerical results are shown in the left inset of Fig.~\ref{fig-entrop}(b) for
different waiting times $t_w$, starting from an infinite temperature initial
condition.
The correlation $C(t_w,t_w+t)$ clearly exhibits aging properties, as the
relaxation time depends strongly on $t_w$.
The same data are shown in the main plot as a function of the rescaled time
$t/t_w$, showing a good collapse at least in the considered time window.
For times $t$ larger than the equilibrium relaxation time $\tau(T)$,
aging becomes interrupted; $\tau(T)$ is seen to increase according to an
Arrhenius law, $\ln \tau(T) \sim 1/T$
--right inset of Fig.~\ref{fig-entrop}(b).

\section{Inherent structures and spatial analysis}
 An inherent structure (IS) is defined as a configuration of the system from
which the global energy cannot be lowered by changing the state of a single
site.
A deterministic algorithm allows to associate unambiguously an IS to any
instantaneous configuration: starting from this configuration, one looks
at each step for the one-site transition that lowers as much as possible
the global energy.
The procedure is iterated until an IS is reached, and the corresponding
configuration $\{q_j^{\textsc{is}}\}$ is recorded.
This algorithm is close in spirit to the zero temperature limit of the
Glauber dynamics, but is still slightly different.
In the Glauber case, all transitions that lower the energy would be treated
as equivalent, and one among them would be chosen randomly, whereas the
procedure used here is purely deterministic.

The basic interpretation of glassy dynamics in phase space is that IS with
lower and lower energies are visited during the aging regime, or at
equilibrium when decreasing temperature.
In a loose sense, IS with lower and lower energies may be considered
as `closer' and `closer' to the ground state.
So from a real space point of view, one might expect these `deep' IS to
minimize the energy at least over some small parts of the system.
To give a well defined meaning to these intuitive ideas, we propose the
following procedure.
Picking up two sites $i$ and $j=i+\xi$, we search for the configuration
$\{\tilde{q}_k\}$ which minimizes the energy upon the interval $[i,j]$,
with the fixed boundary conditions $\tilde{q}_i=q_i^{\textsc{is}}$ and
$\tilde{q}_j=q_j^{\textsc{is}}$.
Then the configuration $\{\tilde{q}_k\}$ is compared with
$\{q_k^{\textsc{is}}\}$ over the interval $[i,j]$.
If both configurations coincide, a larger value $\xi'=\xi+1$
($\xi=2$ in the initial step)
can then be tested, changing $j$ and keeping $i$ fixed.
The procedure is iterated until the largest interval over which
$\{\tilde{q}_k\}$ and $\{q_k^{\textsc{is}}\}$ coincide is found;
the corresponding value of $\xi$ is denoted as $\xi_i$.
Repeating the procedure for all sites $i$, a set of domains
$D_i=[i,i+\xi_i]$ ($i=1, \ldots N$) is obtained. Yet, some of these
domains are actually included in a larger one, since it may happen that
$i+\xi_i = (i+1)+\xi_{i+1}$: in this case, $D_{i+1}$ is included in $D_i$.
Keeping only the intervals $D_i$ that are not included in a larger one,
one finds a non trivial decomposition into almost non overlapping
space intervals, over which the IS considered minimizes the energy.

\begin{figure}
\includegraphics[width=65mm,clip]{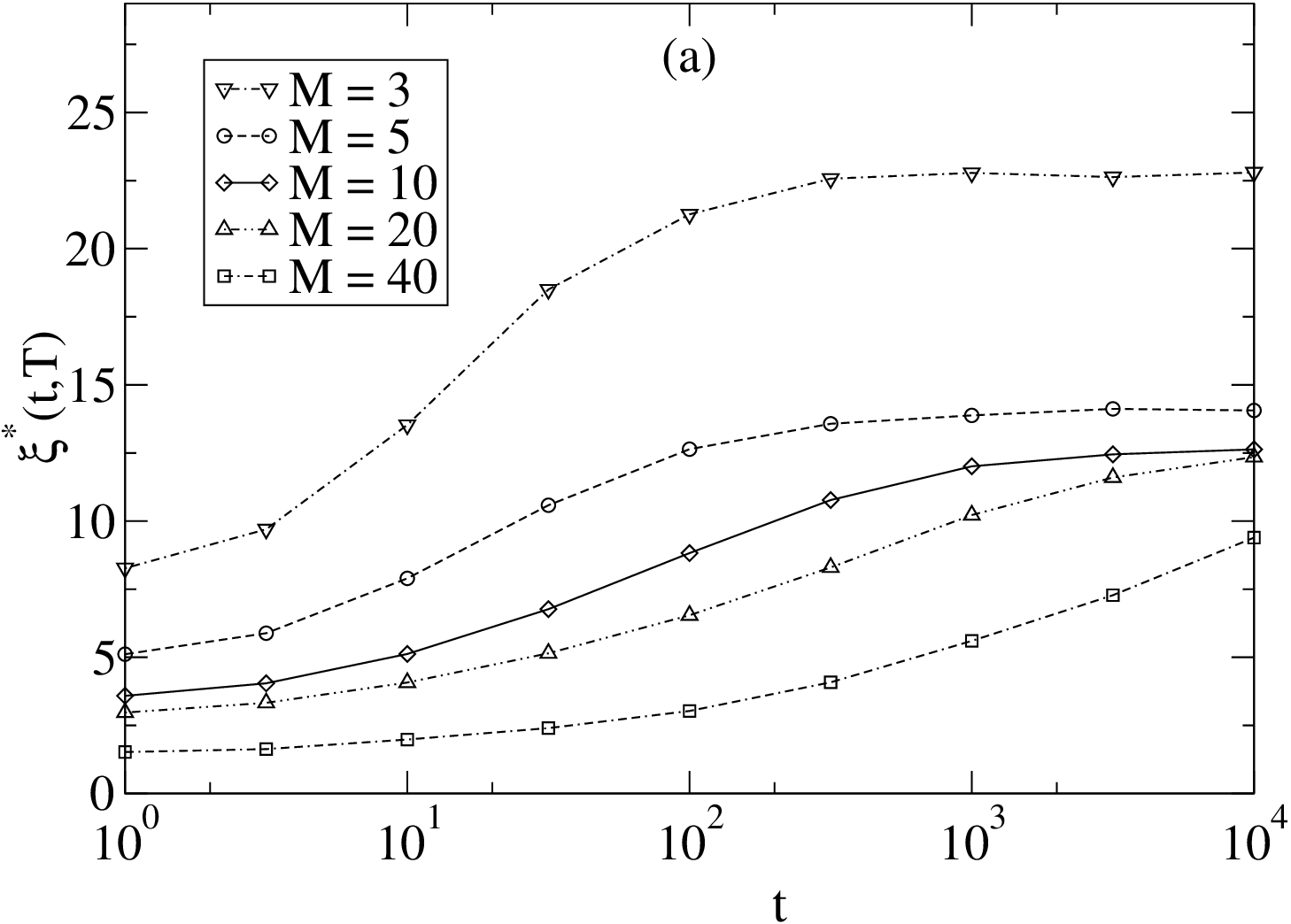} \hfill
\includegraphics[width=65mm,clip]{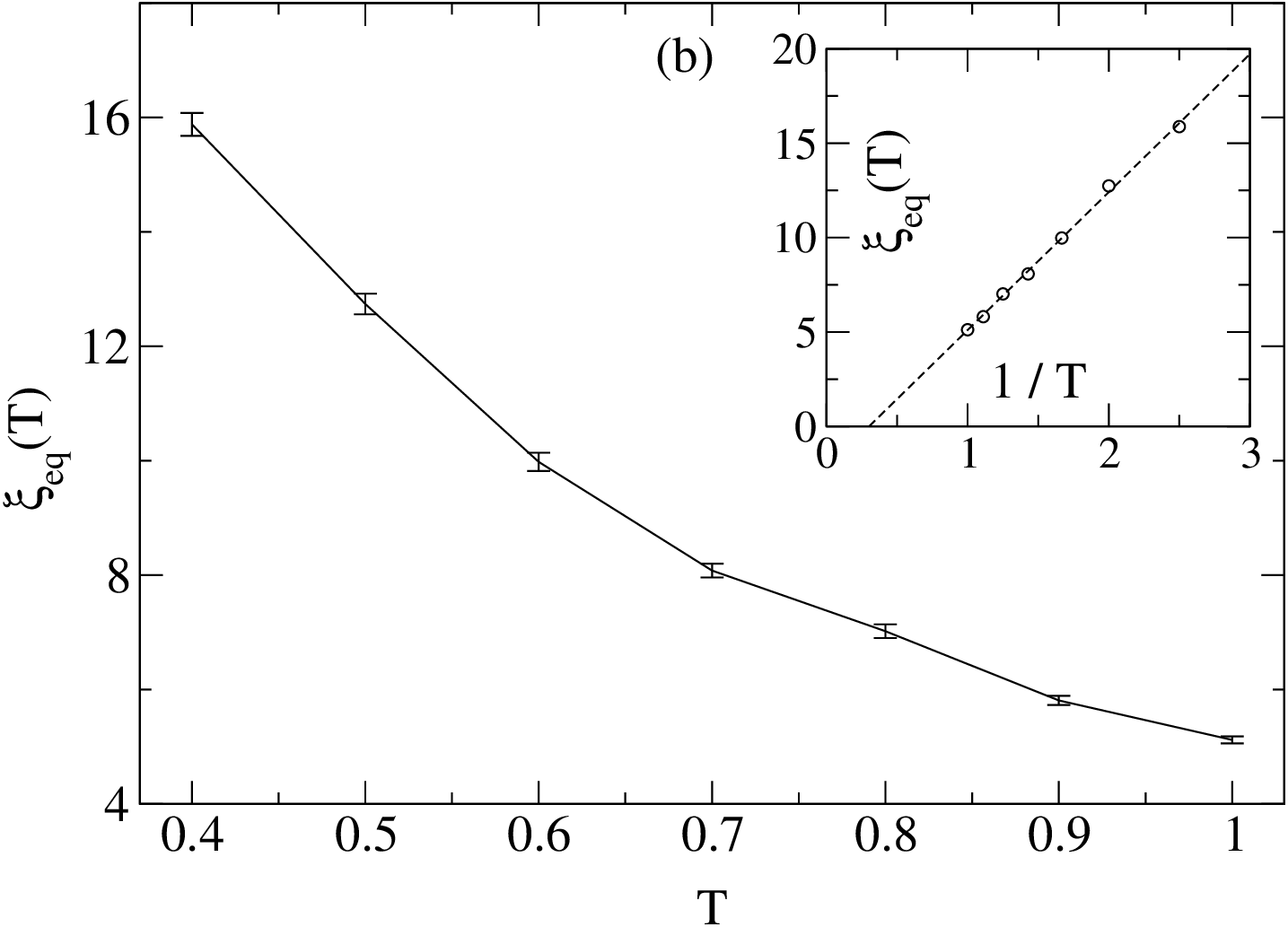}
\caption{\sl (a) $\xi^*(t,T)$ for different values of $M$ ($T=0.5$).
(b) Equilibrium length $\xi_{eq}(T)$ for $M=10$. Inset: same data as a
function of $1/T$, and linear fit $\xi_{eq}(T) \approx a/T-b$ ($a=7.31$,
$b=2.21$).
}
\label{fig-ltT}
\end{figure}

The distribution $P(\xi,t)$ of the size of these regions has been computed
numerically for different times $t$ --see Fig.~\ref{fig-Pell}(a)-- and is
seen to be exponential, at least for $\xi \gtrsim 4$.
From this distribution, the characteristic size $\xi^*(t,T)$ at time $t$ and
temperature $T$ can be deduced from the slope of the exponential;
$\xi^*(t,T)$ is plotted on Fig.~\ref{fig-Pell}(b) for different temperatures.
These simulations were done with relatively small systems ($N=100$), since
the computation time grows like $N^2$, and results were averaged over a
large number of realizations (from $10^3$ to $2\times 10^4$).
We have checked over larger systems that finite size effects are small.

As expected intuitively, the characteristic length scale $\xi^*(t,T)$
increases during the aging regime before saturating to its equilibrium
value at large times.
One can see that the shape of the curves depends only weakly on temperature,
once correctly rescaled --see inset of Fig.~\ref{fig-Pell}(b).
The fact that data rescale as a function of $T\ln t$ indicates that activation
effects play an important r\^ole in the model.
The length $\xi^*(t,T)$ also displays important quantitative
variations with the
value of $M$, as seen from Fig.~\ref{fig-ltT}(a). Yet, the qualitative
behavior remains the same for all $M$.
On the other hand,
the equilibrium length $\xi_{eq}(T)$, obtained from the long time limit of
$\xi^*(t,T)$,
increases markedly (as $1/T$) when temperature is lowered
--see Fig.~\ref{fig-ltT}(b).

\begin{figure}
\includegraphics[width=65mm,clip]{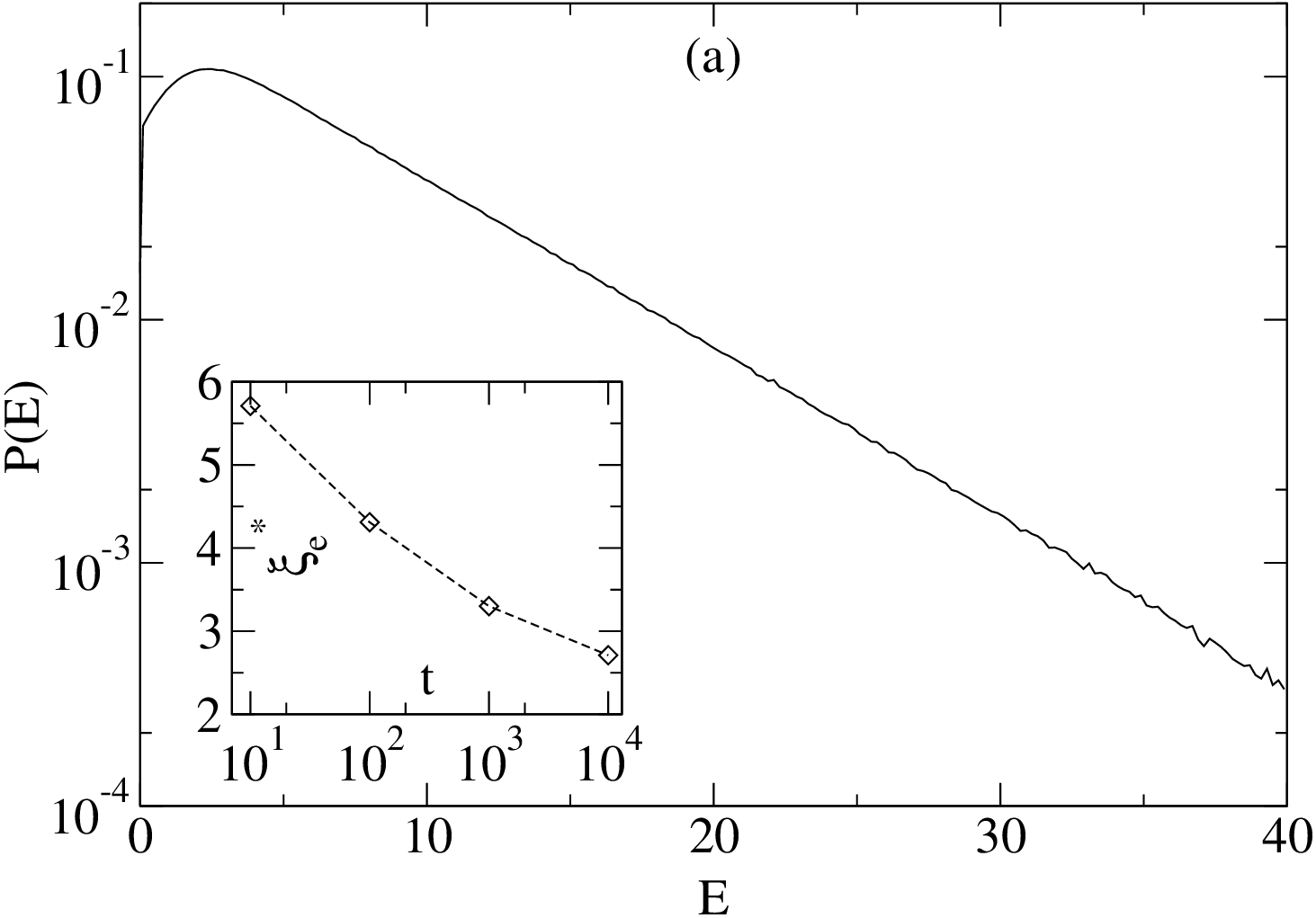} \hfill
\includegraphics[width=65mm,clip]{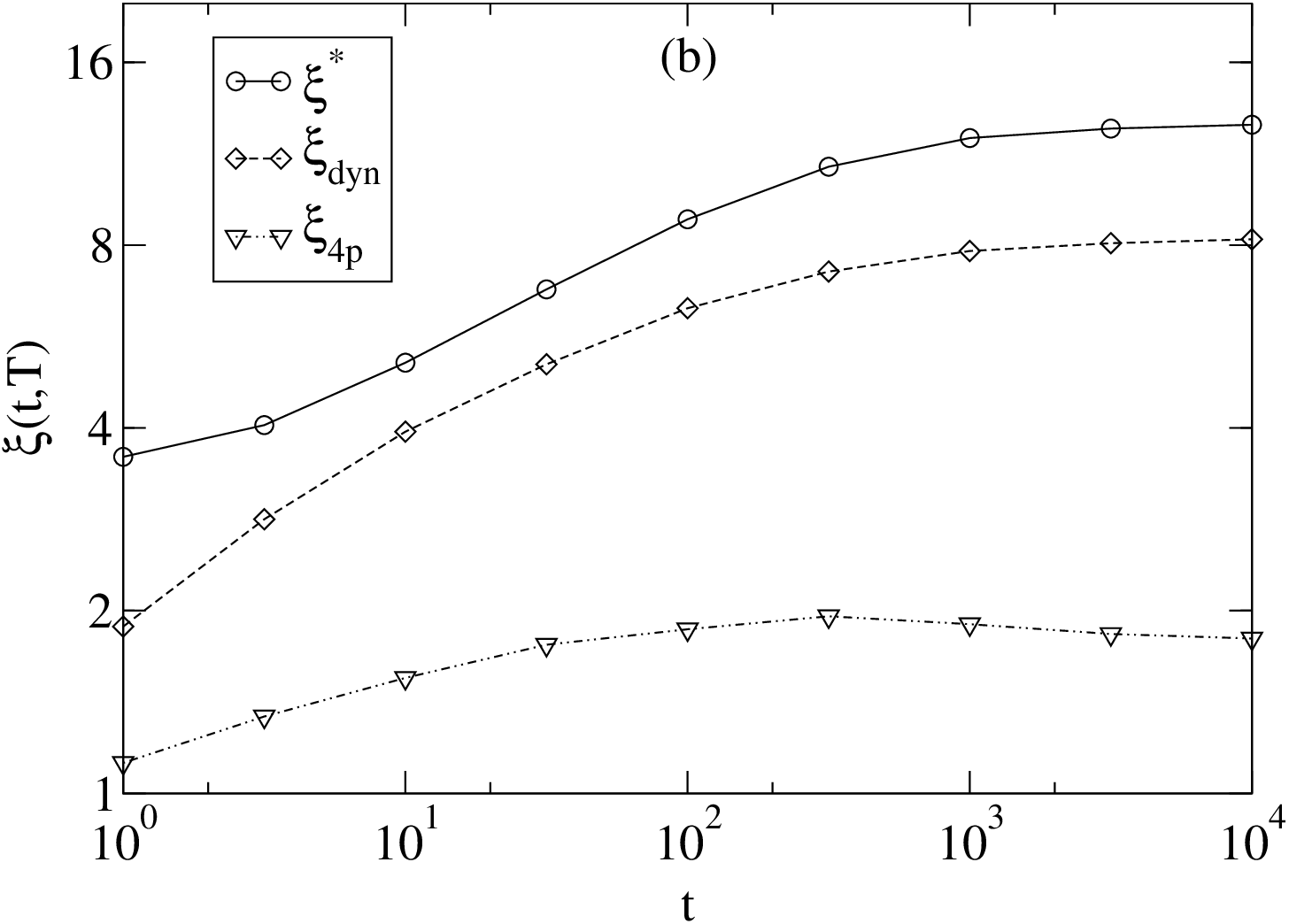}
\caption{\sl (a) A priori distribution $P(E)$ of the energy $E$
(with respect to the ground
state) of the excited domains ($M=10$). Inset: decay of the typical length of
excited domains as a function of time ($T=0.4$, $M=10$).
(b) Comparison between $\xi^*(t,T)$ characterizing the IS,
$\xi_{4p}(t,T)$ obtained from the four-point correlation function,
and $\xi_{dyn}(t,T)$ characterizing the configurations visited dynamically
($T=0.5$, $M=10$).
}
\label{fig-replica}
\end{figure}

One can wonder whether some of the domains found above coincide with the
global ground state of the system.
Interestingly, numerical tests show that this is indeed the case,
and that domains coinciding with the
latter alternate with domains which do not coincide.
Let us denote by $\xi_e^*$ the typical size of these `excited domains',
i.e.~domains that differ from the ground state.
This typical size $\xi_e^*$ need not be the same as the
overall characteristic length
$\xi^*$, since domains coinciding with the ground state may be significantly
larger than the excited ones.
Indeed, numerical simulations show that $\xi_e^*$ {\it decreases}
during the aging regime at fixed temperature, as seen on the inset of
Fig.~\ref{fig-replica}(a), and becomes much smaller than $\xi^*(t,T)$
 --see Fig.~\ref{fig-Pell}(b) for comparison.
So we come up with the following picture for the present model: at large
time after a quench, or in the equilibrium regime,
an IS may be thought of as a set of small excited domains
on top of the ground state.
Moreover, the typical size of these excitations becomes rather small in
comparison with the global average size.

To proceed further, it would be necessary to know the energy distribution
of the excited domains. In order to sample the IS in a uniform way, we
quench the system directly from infinite temperature to determine the IS,
and record the energies (defined with respect to the
ground state) of the excited domains.
The corresponding distribution $P(E)$ is shown in Fig.~\ref{fig-replica}(a),
and is found to be exponential, except for small energies.
We can now test whether this picture is able to make some relevant predictions
concerning the equilibrium length $\xi_{eq}(T)$, and in
particular its $1/T$ behavior.
For simplicity, we assume that the distribution $P(E)$ is purely exponential
(discarding the deviations at small $E$):
$P(E) = \lambda \exp(-\lambda E)$, with $\lambda=0.159$ from the numerical
data.
In equilibrium at temperature $T$, the concentration $c(E)$ of excited
domains with energy $E$ can be estimated as $c(E) = 1/[1+\exp(E/T)]$
(see e.g.~\cite{Berthier-JPG}).
The global concentration $\overline{c}$ of excited domains is then computed
as the average of $c(E)$ over the distribution $P(E)$:
\be
\overline{c} = \int_0^{\infty} dE \, P(E) \, c(E)
= \int_0^{\infty} dE \, \frac{\lambda \, e^{-\lambda E}}{1+e^{E/T}}
\ee
Evaluating the last integral for $T \to 0$ yields
$\overline{c} \approx \lambda T \ln 2$.
The typical distance between these (small size) excited domains is given by
$\overline{\xi} = 1/\overline{c} = \tilde{a}/T$,
with $\tilde{a}=1/(\lambda \ln 2)$.
This distance is precisely the size of the large domains (i.e.~those that
coincide with the ground state), which actually dominate the distribution
$P(\xi,t)$ of domain sizes. Thus this simple argument is indeed able
to predict the correct $1/T$
behavior of $\xi_{eq}(T)$ at low temperature.
The predicted slope $\tilde{a}=9.07$ is in rather good agreement with the
slope $a=7.31$ found from Fig.~\ref{fig-ltT}(b) given the crude
approximations made in the above argument.

\section{Discussion}

A similar decomposition of the IS into domains of minimum energy already
appears in the study of the disordered Ising chain \cite{DerridaGardner},
but with emphasis on the computation of configurational entropy rather than
on the notion of characteristic length scale.
Indeed, in the disordered Ising chain, the introduction of the notion of
`weak links' \cite{DerridaGardner} (i.e.~links that may be either frustrated
or not, in an IS) leads straightforwardly to the identification of
domains upon which the energy is minimized, in a way quite similar to what
happens in the present model.
Yet, to get closer to the present situation, one should rather consider the
disordered Ising chain in the presence of a small external field,
in order to avoid the degeneracy of the ground state.
One then expects to find some excitations on top of the ground state, just
as in the present model.

An open question is to know whether the length scale $\xi^*(t,T)$
characterizing the IS is the same as the length scale $\xi_{4p}$ found from
the usual (two-replica) four-point correlation function. In other words,
are the instantaneous configurations characterized by the same length scale
as the IS?
To check this point, we have computed the four-point correlation function
$C_{ab}(r,t)$ associated to two copies $a$ and $b$ of the system,
with the same disorder, and defined as:
\be
C_{ab}(r,t) = \frac{ \langle S_i(t) S_{i+r}(t) \rangle - \langle S_i(t)
\rangle^2}{\langle S_i(t)^2 \rangle - \langle S_i(t) \rangle^2}
\ee
where $S_i(t)$ is a short notation for $\delta_{q_i^a(t),q_i^b(t)}$.
The length scale $\xi_{4p}$ characterizing the decay of $C_{ab}(r,t)$ is
plotted on Fig.~\ref{fig-replica}(b) as a function of time, for $T=0.5$.
One sees that it remains quite small, of the order of one or two lattice
spacings.

Still, the situation is actually
complicated by the fact that the quantities used
to measure the length scales are different (correlation function versus
distribution of domain sizes), which induces some systematic bias.
Indeed, one can apply the procedure of local energy minimization defined
earlier in this paper for the IS, to the {\it instantaneous configurations}
visited dynamically by the system, without looking to the associated IS.
Interestingly, the distribution of domains found in this way still has an
exponential shape; the corresponding characteristic length scale $\xi_{dyn}$
is plotted on Fig.~\ref{fig-replica}(b). Interestingly, $\xi_{dyn}$ appears
to be much closer to $\xi^*$ than $\xi_{4p}$. When lowering the temperature,
$\xi_{dyn}$ and $\xi^*$ become even closer --data not shown. This may not be
surprising if one considers that IS are obtained from instantaneous
configurations by (partially) removing the thermal noise.
As a result, it seems that IS and instantaneous configurations are essentially
characterized, at least in the present model, by the same length scales
--although they look quite different when comparing $\xi^*$ and $\xi_{4p}$.

Nevertheless, one may expect these two length scales to differ in some
models, as can be seen from the (non-disordered) zero field Ising model.
In this model, the IS are either the two ground states, or
more complicated structures\footnote{If the IS are
determined by a quench with a zero
temperature Glauber dynamics, then non trivial IS appear in the Ising model
in dimension two and above; in two dimensions, these IS are made of
alternating stripes of up and down spins \cite{Ising}.} \cite{Ising},
but they are in any case characterized by a length scale of the order
of the system size.
On the contrary, the dynamical configurations are characterized by a finite
correlation length in equilibrium, and a finite (although growing with time)
typical domain size in the coarsening regime.
Thus these two length scales are indeed different in the Ising model
--although this is a limiting case where one is infinite
in the thermodynamic limit.
So it would be very interesting to see if there exists some models for which
both lengths would be finite, while evolving in independent ways.

\section{Conclusions}

To conclude, we have introduced an analysis of IS in terms of domains upon
which energy is minimized.
The size of these domains is distributed exponentially, and the average size
grows with time at fixed temperature.
The equilibrium length, reached at large times, increases as $1/T$ when
lowering temperature $T$, which can be interpreted within a simple
picture of excited domains.
Besides, this method gives more information than simply comparing the IS
with the ground state, since it shows that domains that do not coincide
with the fundamental still minimize locally the energy.

These results call for future work in order to extend the present procedure
to higher space dimensions. Although this extension may not be trivial, it
should be an important step toward the understanding of the real space
properties of IS.
In addition, it would be very interesting to investigate in different models
the relation between the length scale characterizing the IS and 
the one resulting from the two-replica correlation function.
Either these two length scales should be different at least in some specific
models, or the reason for their identity should be understood.

\acknowledgments The author wishes to thank L. Berthier for important comments, as well as M. Droz for a critical reading of the manuscript.
Numerous fruitful discussions with J.-P. Bouchaud over the last four years are also gratefully acknowledged.

\end{document}